\documentclass[aip,reprint,jcp,graphicx]{revtex4-1}
\usepackage{comment}
\includecomment{reprint} 
\excludecomment{preprint}
\usepackage{amsmath}
\usepackage{amsthm}
\usepackage{amssymb}
\usepackage{graphicx}
\usepackage{microtype}
\usepackage[colorlinks=true]{hyperref}
\hypersetup{
    bookmarks=false,         
    unicode=false,          
    pdftoolbar=false,        
    pdfmenubar=false,        
    pdffitwindow=false,     
    pdfstartview={FitH},    
    pdftitle={},    
    pdfauthor={},     
    pdfsubject={},   
    pdfcreator={},   
    pdfproducer={-}, 
    pdfkeywords={}, 
    pdfnewwindow=true,      
    colorlinks=true,       
    linkcolor=blue,          
    anchorcolor = blue,
    citecolor=blue,        
    filecolor=blue,      
    urlcolor=blue           
}

\usepackage{bm}
\usepackage{color}

\newcommand{\rvec}
{\ensuremath{\mathbf{r}}}
\newcommand{\london}
{L\textsc{ondon }}
\newcommand{\jvec}
{\ensuremath{\mathbf{j}}}

\draft 

\begin{document}

\title{Current Density-Functional Theory using meta-Generalized Gradient Exchange--Correlation Functionals}

\author{James W. Furness}
\email{pcxjf1@nottingham.ac.uk}
\affiliation{School of Chemistry, University of Nottingham, University Park, Nottingham, NG7 2RD, UK}
\author{Joachim Verbeke}
\affiliation{School of Chemistry, University of Nottingham, University Park, Nottingham, NG7 2RD, UK}
\author{Erik I. Tellgren}
\affiliation{Centre for Theoretical and Computational Chemistry, Department of Chemistry, University of Oslo, P. O. Box 1033, Blindern, N-0315, Oslo, Norway}
\author{Stella Stopkowicz}
\affiliation{Centre for Theoretical and Computational Chemistry, Department of Chemistry, University of Oslo, P. O. Box 1033, Blindern, N-0315, Oslo, Norway}
\author{Ulf Ekstr\"om}
\affiliation{Centre for Theoretical and Computational Chemistry, Department of Chemistry, University of Oslo, P. O. Box 1033, Blindern, N-0315, Oslo, Norway}
\author{Trygve Helgaker}
\affiliation{Centre for Theoretical and Computational Chemistry, Department of Chemistry, University of Oslo, P. O. Box 1033, Blindern, N-0315, Oslo, Norway}
\author{Andrew M. Teale}
\email{andrew.teale@nottingham.ac.uk}
\affiliation{School of Chemistry, University of Nottingham, University Park, Nottingham, NG7 2RD, UK}
\affiliation{Centre for Theoretical and Computational Chemistry, Department of Chemistry, University of Oslo, P. O. Box 1033, Blindern, N-0315, Oslo, Norway}

\keywords{Current Density-Functional Theory, Coupled Cluster Theory, Magnetic Properties, Paramagnetic Bonding}
\date{\today}

\begin{abstract}
We present the self-consistent implementation of current-dependent (hybrid) meta generalized gradient approximation (mGGA) density functionals using London atomic orbitals. A previously proposed generalized kinetic energy density is utilized to implement mGGAs in the framework of Kohn--Sham current density-functional theory (KS-CDFT). A unique feature of the non-perturbative implementation of these functionals is the ability to seamlessly explore a wide range of magnetic fields up to 1 a.u. ($\sim 235000$T) in strength. CDFT functionals based on the TPSS and B98 forms are investigated and their performance is assessed by comparison with accurate CCSD(T) data. In the weak field regime magnetic properties such as magnetizabilities and NMR shielding constants show modest but systematic improvements over GGA functionals. However, in strong field regime the mGGA based forms lead to a significantly improved description of the recently proposed perpendicular paramagnetic bonding mechanism, comparing well with CCSD(T) data. In contrast to functionals based on the vorticity these forms are found to be numerically stable and their accuracy at high field suggests the extension of mGGAs to CDFT via the generalized kinetic energy density should provide a useful starting point for further development of CDFT approximations.     
\end{abstract}

\maketitle

\section{Introduction}

The foundations of current density-functional theory (CDFT) and its Kohn--Sham (KS) implementation were established in the late 1980's with the seminal works of Vignale, Rasolt and Geldart~\cite{Vignale1987,Vignale1988,Vignale1988b}, where it was recognised that the exchange--correlation functionals must depend not only on the electronic density, $\rho$, but also the paramagnetic current density $\mathbf{j}_p$ in the presence of an electromagnetic field. Since these early works a large number of theoretical investigations of CDFT have been presented. The foundations of CDFT have sometimes been viewed as controversial. Most recently, Pan and Sahni~\cite{Pan2010,Pan2012,Pan2013} suggested that the physical current density $\mathbf{j}$, rather than $\mathbf{j}_p$, aught to be the fundamental variable in a CDFT and attempted to establish a Hohenberg--Kohn like theorem for this physically appealing alternative choice of variable. Unfortunately, the derivation of this theorem has been shown to be in error~\cite{Tellgren2012,Vignale2013}. Furthermore, the work of Tellgren \emph{et al.}~\cite{Tellgren2012} showed how CDFT may be brought into Lieb's convex-conjugate formulation of DFT~\cite{Lieb1983}, further strengthening its foundations and lending key insight into the more complex relationship between the key densities and potentials in the theory. In particular, it is highlighted that the lack of a Hohenberg--Kohn theorem is not an impediment to a viable CDFT. Recent theoretical works by Lieb and Schrader~\cite{Lieb2013} and Tellgren \emph{et al.}~\cite{Tellgren2014b} have also addressed the issue of $N$-representability in CDFT.  

Despite the theoretical progress in CDFT very few practical implementations of theory have been presented. Most practical studies have either presented calculations based on fixed densities (typically computed at the Hartree--Fock or standard Kohn-Sham level), or have attempted to include CDFT contributions in linear-response calculations. In the context of response theory, implementations have been presented for magnetic properties by Lee \emph{et al.}~\cite{Lee1995} and for excitation energies at the meta generalized gradient approximation (mGGA) level by Bates and Furche~\cite{Bates2012}. Very few fully self-consistent implementations of CDFT capable of treating systems beyond the linear-response regime have been presented, in fact we are only aware of the work by Pittalis \emph{et al.}~\cite{Pittalis1,Pittalis2} in the context of the optimized effective potential method, Zhu and Trickey~\cite{Zhu2014} for atomic systems and our own implementation~\cite{Tellgren2014} for general atomic and molecular species. 

A number of challenges arise when implementing quantum chemical methods for molecules in magnetic fields, the {\sc London} program~\cite{Tellgren2008} has been specifically designed to address these and is utilized throughout the present work. In particular, London atomic orbitals~\cite{LondonAO,Ditchfield1,Ditchfield2} are employed to ensure gauge origin invariant results. For CDFT additional challenges arise since new forms are required for the exchange--correlation functional. Relatively few practical forms for CDFT functionals have been suggested in the literature.

In the present work we will examine the use of mGGAs and hybrid mGGAs for the exchange--correlation energy in the presence of magnetic fields. Functionals in this class depend on the orbital dependent kinetic energy density in the absence of a magnetic field. However, as has been noted in the literature~\cite{Maximoff2004, Tao2005a, Bates2012}, this key quantity is not gauge invariant and so some modification is required for use in a magnetic field. One approach is to replace the kinetic energy density by a generalized form including the paramagnetic current density. This quantity naturally arises in the expansion of the spherically averaged exchange hole, as derived by Dobson~\cite{Dobson1993}. Becke has already suggested the use of this approach to produce a current dependent generalisation of the Becke--Roussel~\cite{Becke1996a} functional. Recently, Bates and Furche~\cite{Bates2012} have also explored a similar generalization of the Tao-Perdew-Staroverov-Scuseria (TPSS) functional~\cite{Tao2003} to calculate excitation energies via response theory. 

We will consider current dependent extensions of the B98~\cite{Becke1998}, TPSS~\cite{Tao2003} and TPSS(h)~\cite{Staroverov2003a} functionals. The use of a modified current-dependent kineitc energy density is denoted by a prefix c throughout the remainder of this work. The non-perturbative nature of the implementation in the {\sc London} program will allow for testing of these functionals in both weak and strong field regimes. The availability of accurate \textit{ab initio} methodologies in the {\sc London} program provides a unique opportunity for the assessment and testing of CDFT functionals at field strengths upto $1$ a.u. In the present work we make use of the recent implementation by Stopkowicz \emph{et al.}~\cite{CC_IMPLEMENTATION} of coupled-cluster (CC) methods with single, double and perturbative triple excitations [CCSD(T)] for benchmarking the CDFT approximations.   

In Section~\ref{sec:theory} we review the simple generalization of mGGA functionals to the CDFT framework due to Dobson and Becke, details specific to the functionals considered in this work are collected in the appendix. In Section~\ref{sec:compdet} we outline some computational details of our calculations. Section~\ref{sec:results} summarizes our findings, assessing the quality of the CDFT approximations by comparison with CCSD(T) data; in Section~\ref{sec:res:magprop} we explore the performance of mGGA functionals for calculating molecular properties in the weak field regime accessible via linear response theory; in Section~\ref{sec:res:parabond} the high field regime is explored by considering the recently proposed perpendicular paramagnetic bonding. The interpretation of this bonding mechanism in the KS-CDFT framework is discussed in Section~\ref{sec:KSCDFTpbond}. Finally, concluding remarks and directions for future work are given in Section~\ref{sec:conclusions}. 

\section{Theory}\label{sec:theory}

In this work we consider the calculation of energies and molecular properties in the presence of a static uniform external magnetic field, $\mathbf{B}$, which may be represented in terms of the vector potential
\begin{equation}\label{extA}
\mathbf{A}(\mathbf{r}) = \frac{1}{2} \mathbf{B} \times (\mathbf{r} - \mathbf{R_\text{G}}),
\end{equation}
where $\mathbf{R}_\text{G}$ is an arbitrary gauge origin. The {\sc London} program makes use of London atomic orbitals~\cite{LondonAO,Ditchfield1,Ditchfield2} to ensure that computed energies and molecular properties are invariant with respect to choice of the gauge origin. These basis functions take the form
\begin{equation}
\omega_{\mu}(\mathbf{r}_K, \mathbf{B}, \mathbf{R}_G) = \exp\left[  \frac{\mathrm{i}}{2} \mathbf{B} \times (\mathbf{R}_\text{G} - \mathbf{R}_\text{K}) \cdot \mathbf{r} \right] \chi_\mu(\mathbf{r}_K)
\end{equation}
where $\chi_\mu(\mathbf{r}_K)$ is a standard Gaussian-type orbital centred at position $\mathbf{R}_K$. These perturbation-dependent basis functions are used to expand the Kohn--Sham molecular orbitals. 

The Kohn--Sham approach to density-functional theory can be extended to CDFT by searching for a non-interacting system of electrons with the same charge and current densities as the physical interacting system: $(\rho_\text{s}(\mathbf{r}), \mathbf{j}_\text{p,s}(\mathbf{r})) = (\rho(\mathbf{r}), \mathbf{j}_\text{p}(\mathbf{r}))$. The KS equations can be written as
\begin{equation}
\left[  \frac{1}{2} p^2 + \frac{1}{2} \{ \mathbf{p}, \mathbf{A}_\text{s} \} + u_\text{s} + \mathbf{s} \cdot [\mathbf{\nabla} \times \mathbf{A}_\text{s}]   \right] \varphi_p  = \varepsilon_p \varphi_p
\end{equation}
where the KS potentials $(u_\text{s}, \mathbf{A}_\text{s})$ are defined as
\begin{equation}
u_\text{s} = v_{\text{ext}} + v_{\text{J}} + v_{\text{xc}} + \frac{1}{2} A_\text{s}^2
\end{equation}
and 
\begin{equation}
\mathbf{A}_\text{s} = \mathbf{A}_\text{ext} + \mathbf{A}_\text{xc}.
\end{equation}
The first three terms in the scalar potential $u_\text{s}$ represent the external (ext), Coulomb (J) and exchange--correlation (xc) potentials defined as the functional derivative of the respective energies with respect to the density, as in the usual KS approach. The final contribution to the scalar potential arises from the non-interacting vector potential $\mathbf{A}_\text{s}$. In addition to the vector potential due to the external field $\mathbf{A}_{\text{ext}}$ as defined in Eq.~(\ref{extA}) the KS vector potential contains an exchange--correlation contribution defined as
\begin{equation}
\mathbf{A}_\text{xc} = \frac{\delta E_{\text{xc}}[\rho, \mathbf{j}_p]}{\delta \mathbf{j}_p}.
\end{equation}  

A central question that immediately arises in CDFT is how the exchange--correlation functional must be modified to include current effects. Whilst the paramagnetic current density is a valid quantity on which to base the universal density functional it can also be shown that the exchange--correlation component must be independently gauge-invariant~\cite{Vignale1988}. This places a significant constraint on the manner in which this quantity may enter any approximate CDFT functional. In contrast to standard DFT, relatively few CDFT functionals have been proposed. The majority of these are based on the the vorticity
\begin{equation}
\bm{\nu}(\rvec) = \nabla \times \left ( \frac{\mathbf{j}_{\mathrm{p}}(\rvec)}{\rho(\rvec)} \right ),
\end{equation}
with 
\begin{equation}
\mathbf{A}_{\text{xc}} = \frac{1}{\rho} \mathbf{\nabla} \times \frac{\delta E_\text{xc}[\rho, \bm{\nu}]}{\delta \bm{\nu}}
\end{equation}
as proposed by Vignale, Rasolt and Geldart (VRG)~\cite{Vignale1988b}. The original VRG form for the exchange--correlation energy was parameterized using Monte Carlo simulations of the high density limit\cite{Vignale1987}. A number of re-parameterisations for this form have been suggested based on accurate calculations in the high-density regime~\cite{Lee1995, Orestes2003,Tao2005, Tao2006}. Higuchi and Higuchi~\cite{Higuchi2007} (HH) have also presented a vorticity dependent form, derived to obey known exact relations for the CDFT exchange--correlation functional. 

Whilst the vorticity is a theoretically convenient choice to ensure the gauge invariance of the exchange--correlation energy it has been observed that in practical self-consistent calculations it can lead to significant numerical stability issues~ \cite{Zhu2014, Tellgren2014}. How severe these issues are depends on the exact parameterization of the the functional form, however, in all cases some degree of numerical regularization is required to ensure that the self-consistent field solution of the Kohn--Sham equations can be obtained. Furthermore, molecular properties computed by such calculations display an un-acceptably strong dependence on the regularization parameters -- with no obvious convergence towards a single value. Clearly this raises questions as to how appropriate such forms are for use in quantum chemical calculations.

Most practical mGGAs make use of the kinetic energy density 
\begin{equation}
\label{tau}
\tau_{\sigma} = \sum_i^{\mathrm{occ}}  \nabla \varphi^*_{i\sigma} \cdot \nabla \varphi_{i\sigma} ,
\end{equation}
in their construction, where $\varphi_i$ are the occupied KS orbitals and $\sigma$ is the electron spin index. This term is gauge dependent and as a result an unmodified meta-GGA type functional form cannot be used to describe a system with a non-zero magnetic vector potential. To resolve this issue the gauge independence of the exchange--correlation functional must be restored. A natural modification, which can be applied to any mGGA dependent on the kinetic energy density $\tau(\rvec)$, arises in the work of Dobson\cite{Dobson1992, Dobson1993} who generalized the expansion of the exchange-hole to include the case of non-zero current densities.  

The spherically averaged exchange hole at zero field can be modelled using a Taylor expansion~\cite{Becke1983} and is commonly considered~\cite{Becke1989,Becke1996a,Becke1998} up to the quadratic term
\begin{equation}\label{Q}
Q_{\sigma} = \frac{1}{6}\left [ \nabla^2\rho_{\sigma} - 2\tau_{\sigma} + \frac{(\nabla \rho_{\sigma})^2} { 2\rho_{\sigma}}  \right ].
\end{equation}
This expansion can be generalised to non-zero field and the curvature term becomes~\cite{Dobson1992, Dobson1993,Becke1996a}
\begin{equation}\label{cQ}
Q_{\sigma} = \frac{1}{6}\left [ \nabla^2\rho_{\sigma} - 2\tau_{\sigma} + \frac{(\nabla \rho_{\sigma})^2} { 2\rho_{\sigma}} + \frac{2  \left |\mathbf{j}_{p\sigma} \right |^2}{\rho_\sigma} \right  ],
\end{equation} 
where $\mathbf{j}_{p\sigma}$ is the paramagnetic current density
\begin{equation}
\mathbf{j}_{p\sigma} = - \frac{i}{2} \sum_i^{\text{occ}} \left[ \varphi_{i\sigma}^* \nabla \varphi_{i\sigma} - \varphi_{i\sigma} \nabla \varphi_{i\sigma}^*   \right].
\end{equation}
Comparing Eqs.~(\ref{Q}) and~(\ref{cQ}) it is possible to identify a correction to the conventional $\tau(\rvec)$ that is gauge invariant and may be utilized in mGGA functionals,
\begin{equation}\label{tauGI}
\tau_{\sigma} \rightarrow \tilde \tau_{\sigma} = \tau_{\sigma} - \frac{ \left | \jvec_{\mathrm{p}\sigma} \right | ^2}{\rho_{\sigma}},
\end{equation}

The use of Eq.~(\ref{tauGI}) has been put forward many times in the literature. Becke suggested its use in the Becke--Roussel model~\cite{Becke1996a} to generate a current dependent analogue of this functional. He also suggested that this quantity could be used to define a current dependent inhomogeneity parameter in the more empirical B98 functional~\cite{Becke1998}. It has also been suggested for use to generalize the TPSS functional~\cite{Tao2003} by Tao~\cite{Tao2005a}. Recently Bates and Furche~\cite{Bates2012} considered the application of the resulting cTPSS functional in the calculation of excitation energies via response theory. In the present work we consider the application of mGGA functionals with this modification to calculation magnetic properties in the weak and strong field regimes in a non-perturbative manner. In particular we consider three functional forms, cB98, cTPSS and cTPSS(h), where the prefix c denotes the use of the modified $\tilde \tau_{\sigma}$ in Eq.~(\ref{tauGI}). The Appendix gives some details of these functional forms to show how these modifications enter.

\section{Computational Details}\label{sec:compdet}

Unless otherwise indicated all calculations in this work use the aug-cc-pCVTZ basis set~\cite{Dunning1989,Kendall1992}. All DFT calculations have been performed using the \london\cite{Tellgren2008} program. This code utilizes the XCFun library~\cite{Ekstro2010} for the evaluation of the density functionals and their derivatives. The modifications of Eq.~(\ref{tauGI}) and the functionals cB98 and cTPSS have been added to the XCFun library. In addition we investigate the use of a hybrid form of cTPSS, denoted cTPSS(h), based on the TPSS(h) functional of Ref.~\onlinecite{Staroverov2003}. The quality of the CDFT functionals cB98, cTPSS and cTPSS(h) is assessed by comparison with CCSD(T) data. For comparison Hartree--Fock (HF), local density approximation (LDA), Perdew-Burke-Ernzerhof (PBE)~\cite{Perdew1996}, and Keal-Tozer-3 (KT3)~\cite{Keal2004} density-functional results are also presented. The latter is of particular interest since it is specifically designed for the calculation of nuclear magnetic resonance (NMR) shielding constants. 

The performance of these approximations will be considered in two regimes; the weak field regime accessible by linear response calculations and the strong field regime only accessible via non-perturbative calculations. In the weak field regime we will consider the calculation of molecular magentizabilities and NMR shielding constants for the 26 small molecules in Table~\ref{tab:testset}. Errors for these quantities are presented relative to the benchmark data of Ref.~\onlinecite{Teale2013a}. Results are also compared with those including corrections from the Tao-Perdew parameterization~\cite{Tao2005} of the Vignale-Rasolt-Geldart functional~\cite{Vignale1988}, taken from Ref.~\onlinecite{Tellgren2014}.  In the strong field regime we consider the prediction of perpendicular paramagnetic bonding~\cite{Lange2012} in a field strength of 1 a.u. perpendicular to the internuclear axes of H$_2$, He$_2$, HeNe and Ne$_2$. These non-perturbative calculations are assessed against CCSD(T) results computed using the implementation of Ref.~\onlinecite{CC_IMPLEMENTATION} in the \london program.    
\begin{table}
\caption{The test set of molecules for which accurate benchmark CCSD(T) data from Ref.~\onlinecite{Teale2013a} was available.}
\begin{tabular}{|ccccccc|}\label{tab:testset}
HF & CO & N$_2$ & H$_2$O & HCN & HOF & LiH  \\
NH$_3$ & H$_2$CO & CH$_4$ & C$_2$H$_4$ & AlF & CH$_3$F & C$_3$H$_4$ \\
FCCH & FCN & H$_2$S & HCP & HFCO & H$_2$C$_2$O & LiF \\
 & N$_2$O & OCS &  H$_4$C$_2$O & PN & SO$_2$  & \\
\end{tabular}
\end{table}

\section{Results}\label{sec:results}

\subsection{The weak field regime: magnetic properties}\label{sec:res:magprop}

We commence by considering the molecular magnetizabilities and NMR shielding constants of the 26 small molecules in Table~\ref{tab:testset}. The magnetizability tensor elements, $\xi_{\alpha,\beta}$, are defined as
\begin{equation}
\xi_{\alpha, \beta} = \left. \frac{\partial^2 E(\mathbf{B})}{\partial B_\alpha \partial B_\beta} \right|_{\mathbf{B = 0}}
\end{equation} 
where $\alpha$ and $\beta$ label cartesian components of the tensor and magnetic field. The NMR shielding tensor for a given nucleus $K$ is defined by 
\begin{equation}
\sigma_{K; \alpha, \beta} = \left. \frac{\partial^2 E(\mathbf{B}, \mathbf{M}_K)}{\partial B_\alpha \partial M_{K, \beta}} \right|_{\mathbf{B = 0}, \mathbf{M}_K = \mathbf{0}}
\end{equation}   
where $\mathbf{M}_K$ is the nuclear magnetic moment of nucleus $K$. 
These properties can be accessed non-perturbatively in the \london program by explicit calculation of the energy in the presence of the perturbing fields. Details of this procedure are given in Ref.~\onlinecite{Tellgren2014}, here we compute values for each method considered in the same manner, facilitating a comparison with previous results. 

Given that for many density-functional approximations these singlet second order magnetic response properties can be accessed by standard linear response methods in a variety of programs, this approach may seem cumbersome. However, it should be noted that the implementation of the new CDFT approaches in this framework is much more straightforward, requiring only an implementation of the functional and the derivatives required for construction of the KS matrix. More importantly, as we will see in Section~\ref{sec:res:parabond}, this non-perturbative approach allows us to seamlessly explore the behaviour of new approximations in much stronger fields -- inaccessible via linear response theory. This means that \london provides a powerful testbed for new CDFT functionals.

To quantify the accuracy of the DFT approaches for the calculation of these properties we compare our results with the CCSD(T) benchmark values of Ref.~\onlinecite{Teale2013a}. Specifically, we use the values at the CCSD(T)/aug-cc-pCV[TQ]Z level -- which have been extrapolated to the basis set limit using the procedure of Refs.~\onlinecite{Teale2013a,Lutnaes2009}. 

\begin{table}
\caption{The mean error (ME), mean absolute error (MAE), and standard deviation (SD) of magnetic properties relative to the CCSD(T) benchmark data of Refs.~\cite{Teale2013a,Lutnaes2009}. See also Figs. \ref{fig:MagBox} and \ref{fig:SheBox}.  }\label{tab:errs}
\begin{tabular}{l|ccc|ccc}
\hline
\hline
 & \multicolumn{3}{c|}{Magnetizability} & \multicolumn{3}{c}{NMR Shielding} \\
  & \multicolumn{3}{c|}{(10$^{-30} \mathrm{J T}^{-2}$)} & \multicolumn{3}{c}{(ppm)} \\
 & ME & MAE & SD & ME & MAE & SD \\
\hline
HF & -3.06 & 6.35 & 7.29 & -15.18 & 21.40 & 40.97 \\
LDA & 5.01 & 9.18 & 10.86 & -24.81 & 24.85 & 30.00 \\
PBE & 6.75 & 8.81 & 9.03 & -19.66 & 19.78 & 21.46 \\
PBE+VRG(TP) & 7.85 & 9.61 & 9.44 & -20.33 & 20.46 & 22.43 \\
KT3 & 8.18 & 8.95 & 7.83 & -6.53 & 8.94 & 13.13 \\
KT3+VRG(TP) & 9.18 & 9.83 & 8.19 & -7.45 & 9.18 & 13.37 \\
B97-2 & 5.46 & 5.84 & 5.96 & -16.34 & 16.48 & 20.60 \\
cB98 & 0.52 & 4.84 & 6.58 & -12.44 & 12.66 & 17.79 \\
cTPSS & 7.13 & 7.51 & 6.76 & -14.14 & 14.35 & 15.61 \\
cTPSS(h) & 6.41 & 6.51 & 6.00 & -14.33 & 14.52 & 16.68 \\
\hline
\hline
\end{tabular}
\end{table}

\begin{figure}
\includegraphics[width=\columnwidth]{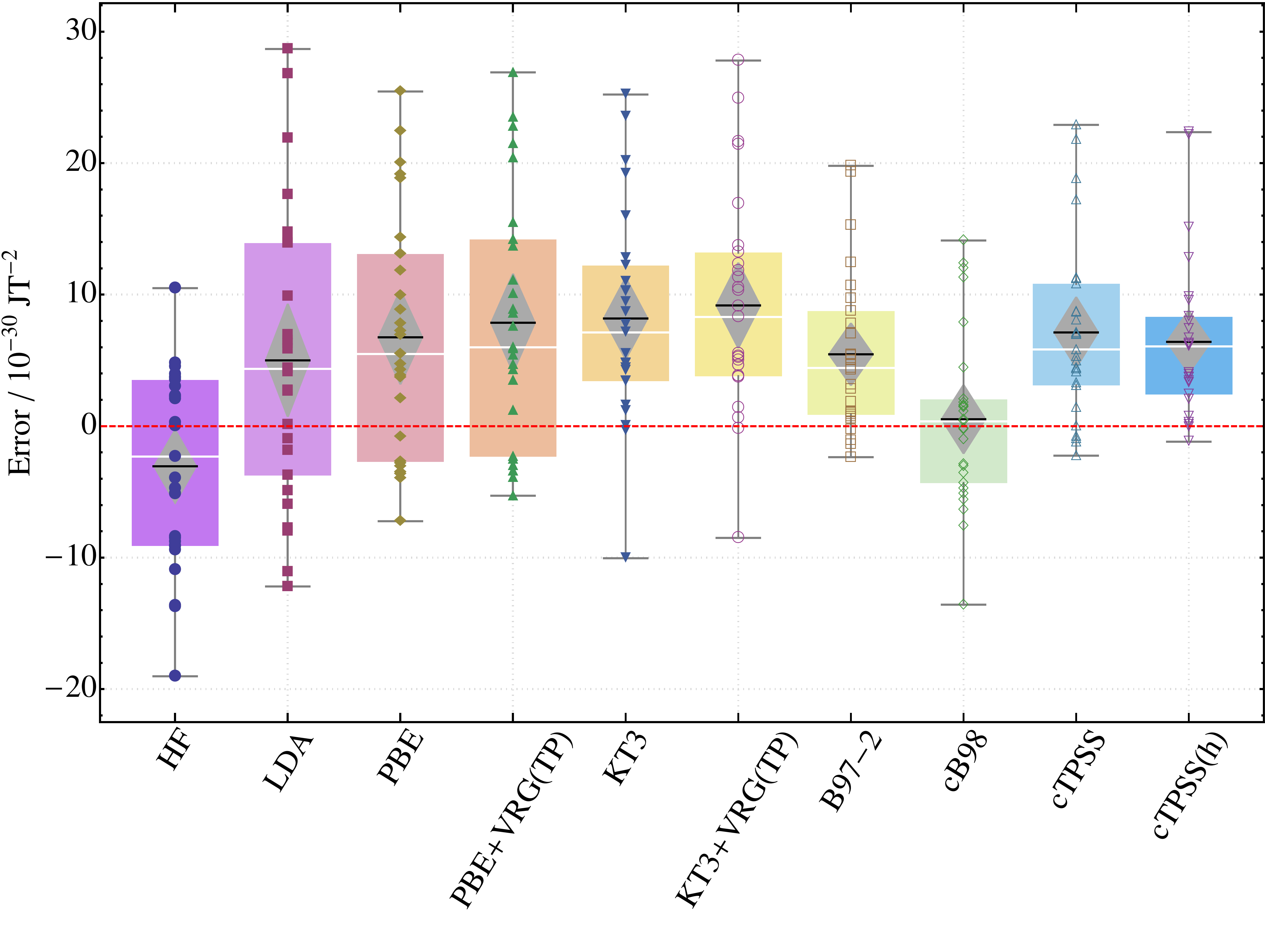}
\caption{Box-whisker plot of errors in (C)DFT molecular magnetizabilities relative to the extrapolated CCSD(T) values of Ref.~\onlinecite{Lutnaes2009}. All calculations use the aug-cc-pCVTZ basis set. See the text for further details. }\label{fig:MagBox}
\end{figure}

\begin{figure}
\includegraphics[width=\columnwidth]{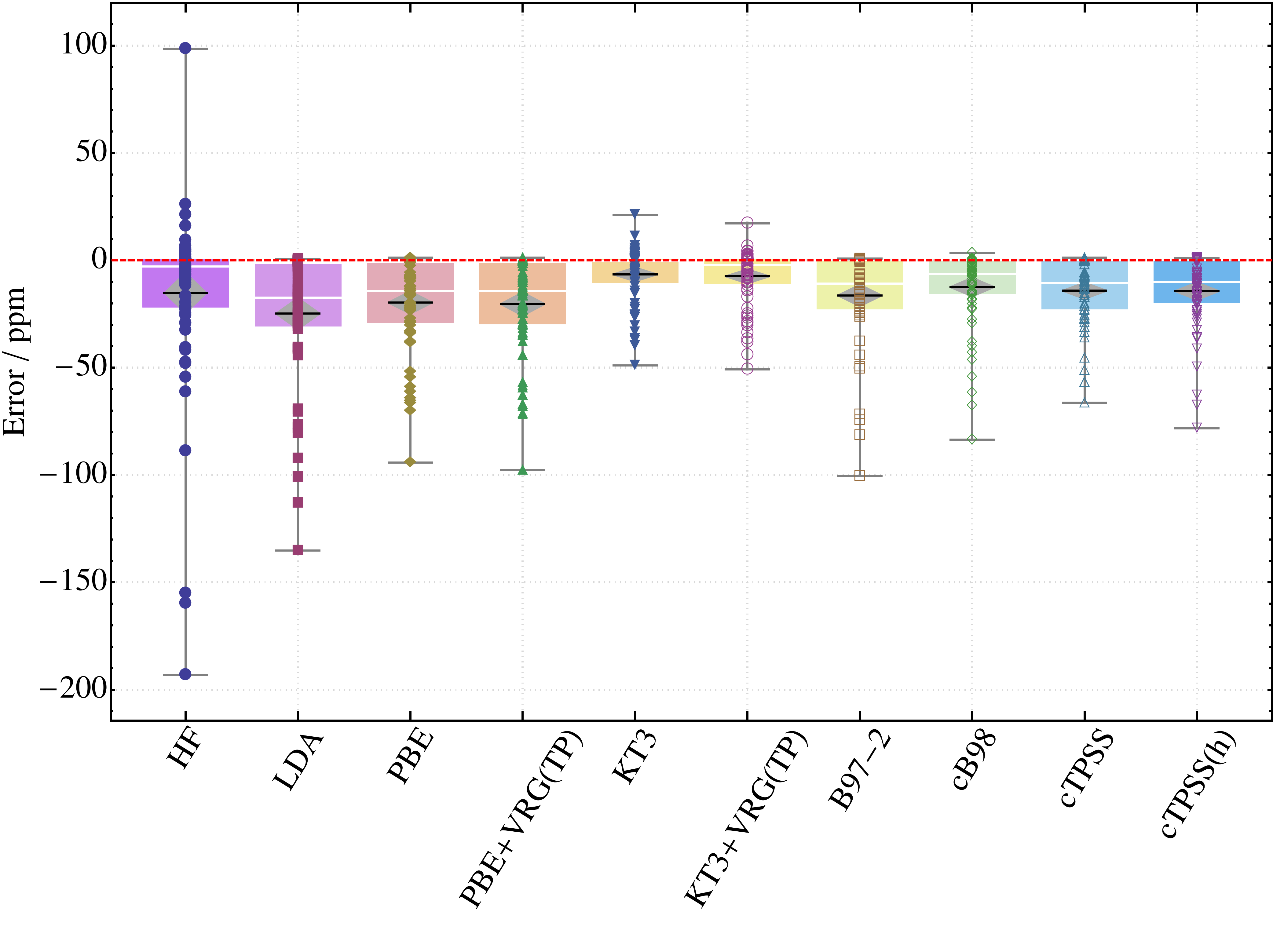}
\caption{Box-whisker plot of errors in (C)DFT NMR shielding constants relative to the extrapolated CCSD(T) values of Ref.~\onlinecite{Teale2013a}. All calculations use the aug-cc-pCVTZ basis set. See the text for further details.}\label{fig:SheBox}
\end{figure}

The errors in the calculated magnetizabilities and NMR shielding constants are summarised in Table \ref{tab:errs} and presented graphically in Figures~\ref{fig:MagBox} and~\ref{fig:SheBox} as box-whisker plots. In these plots individual points represent the errors for each system relative to the reference values, the upper and lower fences of the whiskers denote the maximum positive and negative errors respectively and the coloured boxes enclose errors between the 25\% and 75\% quantiles. Mean and median errors are marked in the plots by horizontal black and white lines, respectively. Grey diamonds are used to represent the confidence intervals.

For the molecular magnetizabilities it is clear from the error measures in Table~\ref{tab:errs} and their representation in Figure~\ref{fig:MagBox} that none of the functionals offers high accuracy. The GGA functionals PBE and KT3 in particular do not offer significant improvements over LDA. Whilst their minimum and maximum errors are slightly improved, the mean errors actually deteriorate. Similar observations were made in Ref.~\onlinecite{Lutnaes2009} for these type of functionals. The B97-2 functional gives slightly reduced errors and this is consistent with previous conclusions~\cite{Lutnaes2009} that for magnetizabilities the inclusion of HF exchange may be beneficial. At the GGA level the underlying functionals are already gauge invariant but do not depend explicitly on the paramagnetic current density. To introduce this dependence the VRG functional may be added. In our earlier work~\cite{Tellgren2014} we found that this correction can be numerically problematic and that the most stable parameterization of this functional to date is that put forward by Tao and Perdew~\cite{Tao2005}, denoted VRG(TP). For comparison we include here the PBE+VRG(TP) and KT3+VRG(TP) results. It is clear that the inclusion of the VRG(TP) correction actually worsens the agreement of the results with CCSD(T). At the mGGA level the inclusion of current is mandatory to ensure the exchange--correlation evaluation is gauge independent. 

Here we investigate the cB98, cTPSS and cTPSS(h) functionals, the TPSS and TPSS(h) forms are similar in performance to PBE -- offering marginal improvements on some error measures, with TPSS(h) performing slightly better than TPSS. The cB98 form gives the best performance of all the functionals considered. It is noteworthy that in the mGGA class the mean errors reduce as more HF exchange is included in the functional -- with cTPSS, cTPSS(h) and cB98 containing 0\%, 10\% and  19.85\% HF exchange respectively. Suggesting that for this property that the treatment of exchange may be the dominant factor in the errors. Since the treatment of long-range exchange is not rectified in the transition from GGA to mGGA type functionals (and only partially corrected by a global admixture), then it may be that this factor far out weighs any improvements due to the inclusion of current effects.      

It is worth emphasizing that in the course of our investigation we found that the implementation of the mGGA functionals including a generalized kinetic energy density was straightforward. In particular we found that no special care was required with respect to numerics compared with standard functionals and that in practical use the functionals are robust and self-consistent calculations using these functionals converge without significant difficulty. This sharply contrasts the behaviour for the VRG functionals as investigated in Refs.~\onlinecite{Tellgren2014, Zhu2014}. 

The results for NMR shielding constants are presented in Figure~\ref{fig:SheBox}. Here we see LDA is poor as expected. In addition KT3, which was designed for these properties, is the best performing functional -- significantly improving over the standard PBE GGA functional and the B97-2 hybrid functional. In this case we see the addition of the VRG(TP) correction to PBE and KT3 has little effect, very slightly deteriorating the results. The mGGA results for cB98, cTPSS and cTPSS(h) are intermediate between PBE and KT3 -- offering small systematic improvements over PBE. Again B98 produces the best results of the mGGA functionals. 

On the whole the quality of the mGGA results at modest field strengths may be regarded as disappointing. The overall errors suggest that mGGAs may offer modest improvements over conventional GGA functionals such as PBE -- though they cannot compete with GGAs tailored to specific properties. This is broadly consistent with findings by Bates and Furche for the calculation of excitation energies using cTPSS~\cite{Bates2012}. Since current corrections are known to be relatively small it is important that the underlying functional should be relatively accurate. For the mGGAs considered here there are known weaknesses (for example in the treatment of long-range exchange) that may obscure the effect of the current dependence. For the case of NMR shielding constants a more detailed analysis of the significance of current effects and how these interplay with errors in a range of density functionals is presented in Ref.~\onlinecite{SarahPCCP}. We will now examine how these functionals perform when the magnetic field becomes much higher and has a stronger effect on the electronic structure. 

\subsection{The strong field regime: paramagnetic bonding}\label{sec:res:parabond}

One approach to explore whether or not the inclusion of current effects via the modified kinetic energy density of Eq.~(\ref{tauGI}) is physically reasonable is to increase the strength of the magnetic field. Lange et. al.\cite{Lange2012} have recently performed full configuration-interaction (FCI) calculations at high field that have uncovered a new mechanism for chemical bonding in the presence of a strong magnetic field. This new bonding has been termed perpendicular paramagnetic bonding and occurs at field strengths similar to those found on some white dwarf stars. Since this work Murdin \emph{et al.}~\cite{Murdin2013} have shown that phosphorus and selenium doped silicon semiconductors can produce a viable laboratory analogue of free hydrogen~\cite{Murdin2013} and helium~\cite{Litvinenko2014} in strong magnetic fields. The description of these types of systems via quantum chemistry will require less computationally demanding approaches -- and CDFT is one strong candidate for the simulation of these systems.  

\begin{figure}[h] 
\includegraphics[width=\columnwidth]{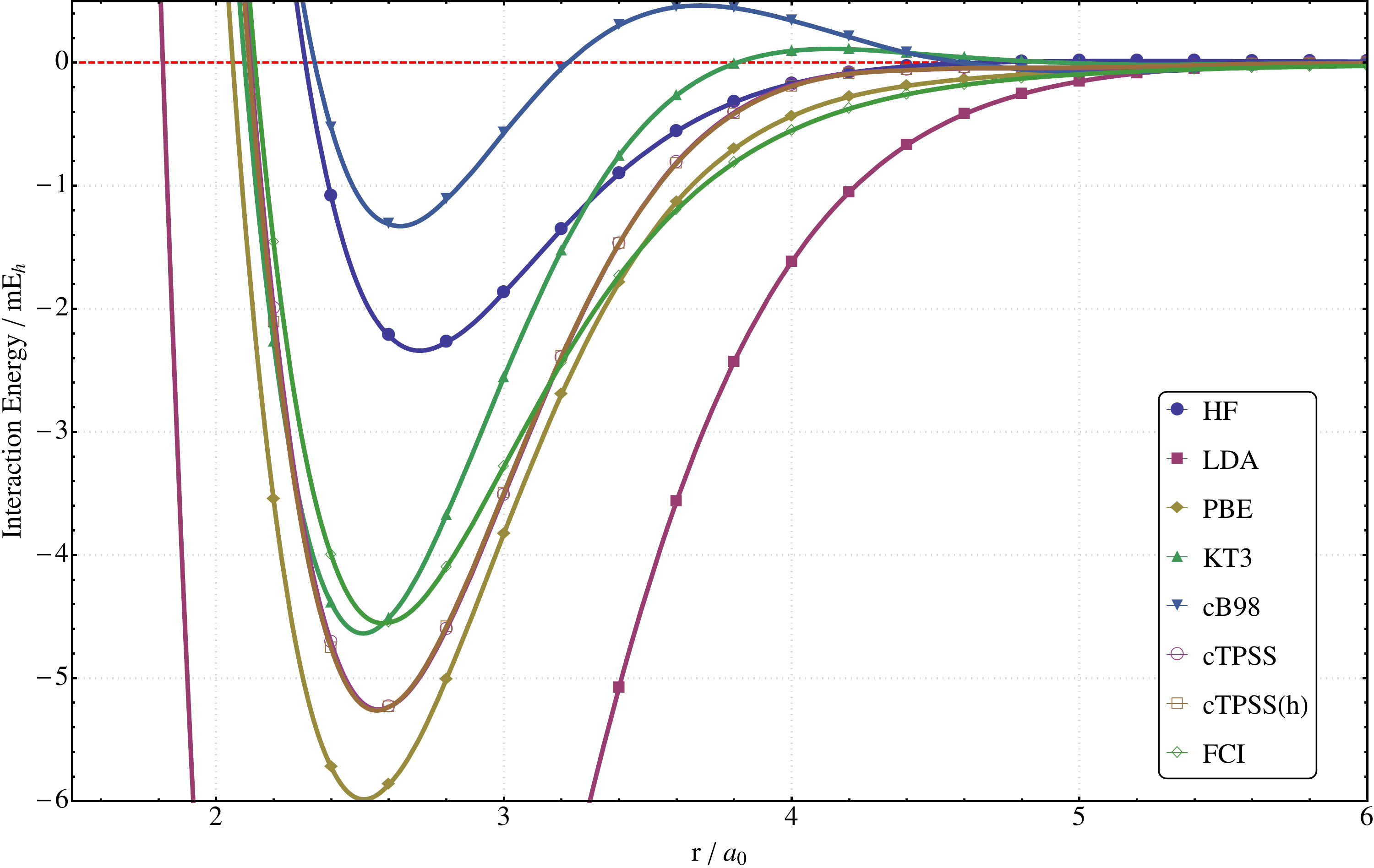}
\caption{Potential energy curve for the $^3\Sigma_u^+ (1 \sigma_g 1 \sigma_u^*)$ state of the H$_2$ molecule in a magnetic field of 1 a.u. perpendicular to the bonding axis for a variety of methods with the aug-cc-pCVTZ basis set.}\label{fig:h2}
\end{figure}

\begin{figure}[h] 
\includegraphics[width=\columnwidth]{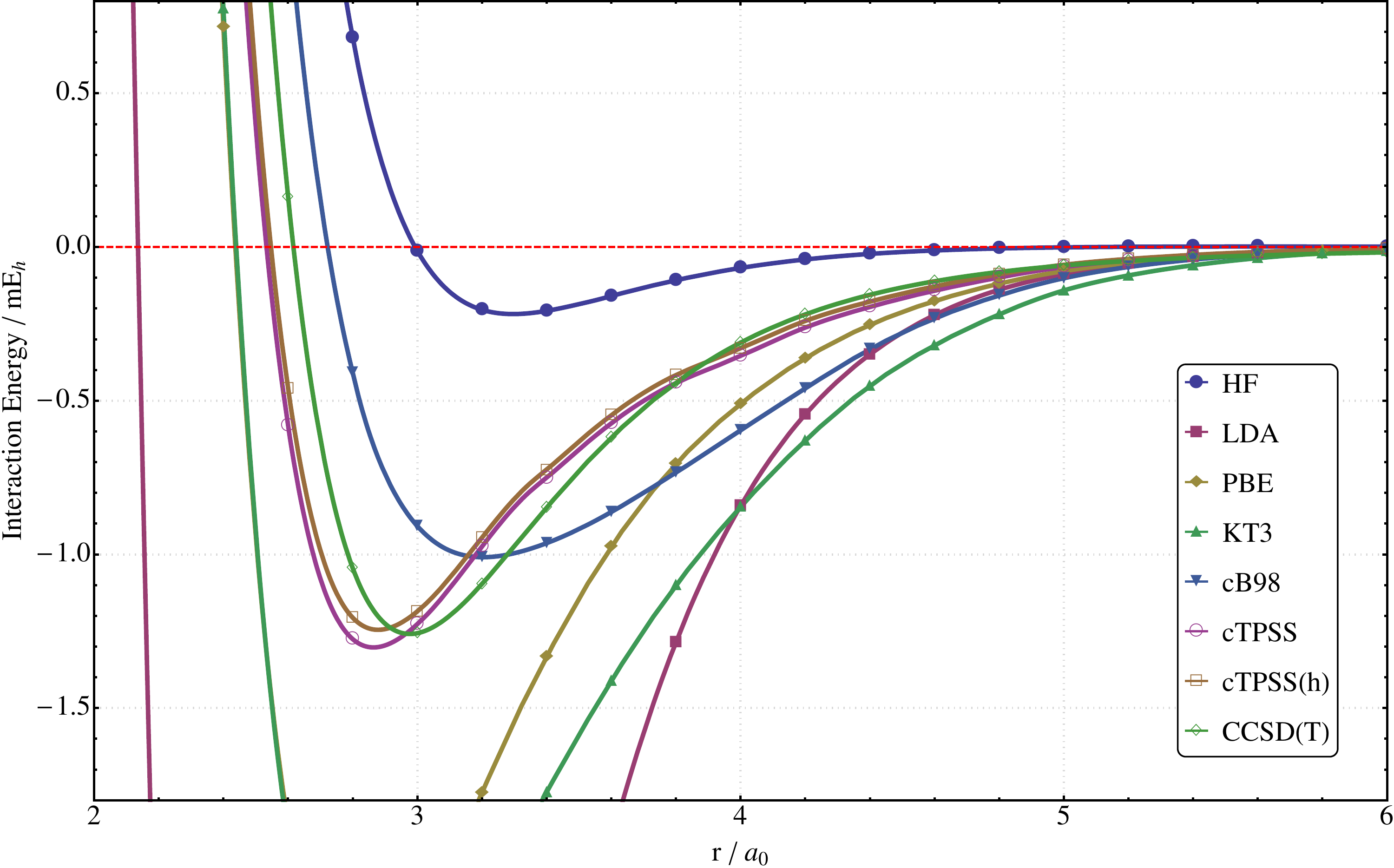}
\caption{Potential energy curve for the lowest $^1\Sigma_g^+$ state of He$_2$ in a magnetic field of 1 a.u. perpendicular to the bonding axis for a variety of methods with the aug-cc-pCVTZ basis set.}\label{fig:hehe}
\end{figure}

\begin{figure}[h]
\includegraphics[width=\columnwidth]{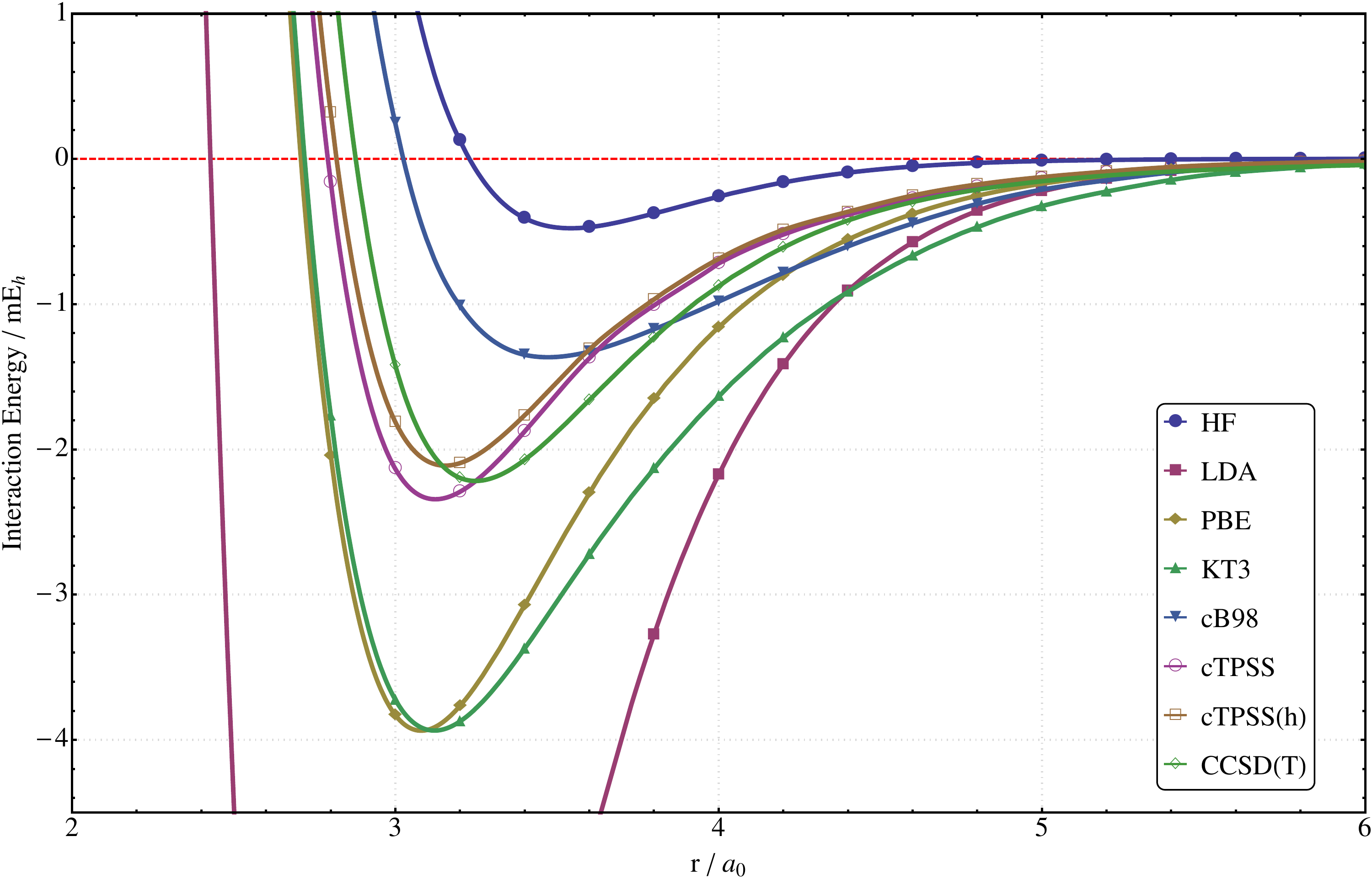}
\caption{Potential energy curve for the lowest $^1\Sigma_g^+$ state of the NeHe dimer in a magnetic field of 1 a.u. perpendicular to the bonding axis for a variety of methods with the aug-cc-pCVTZ basis set. }\label{fig:nehe}
\end{figure}

\begin{figure}[h]
\includegraphics[width=\columnwidth]{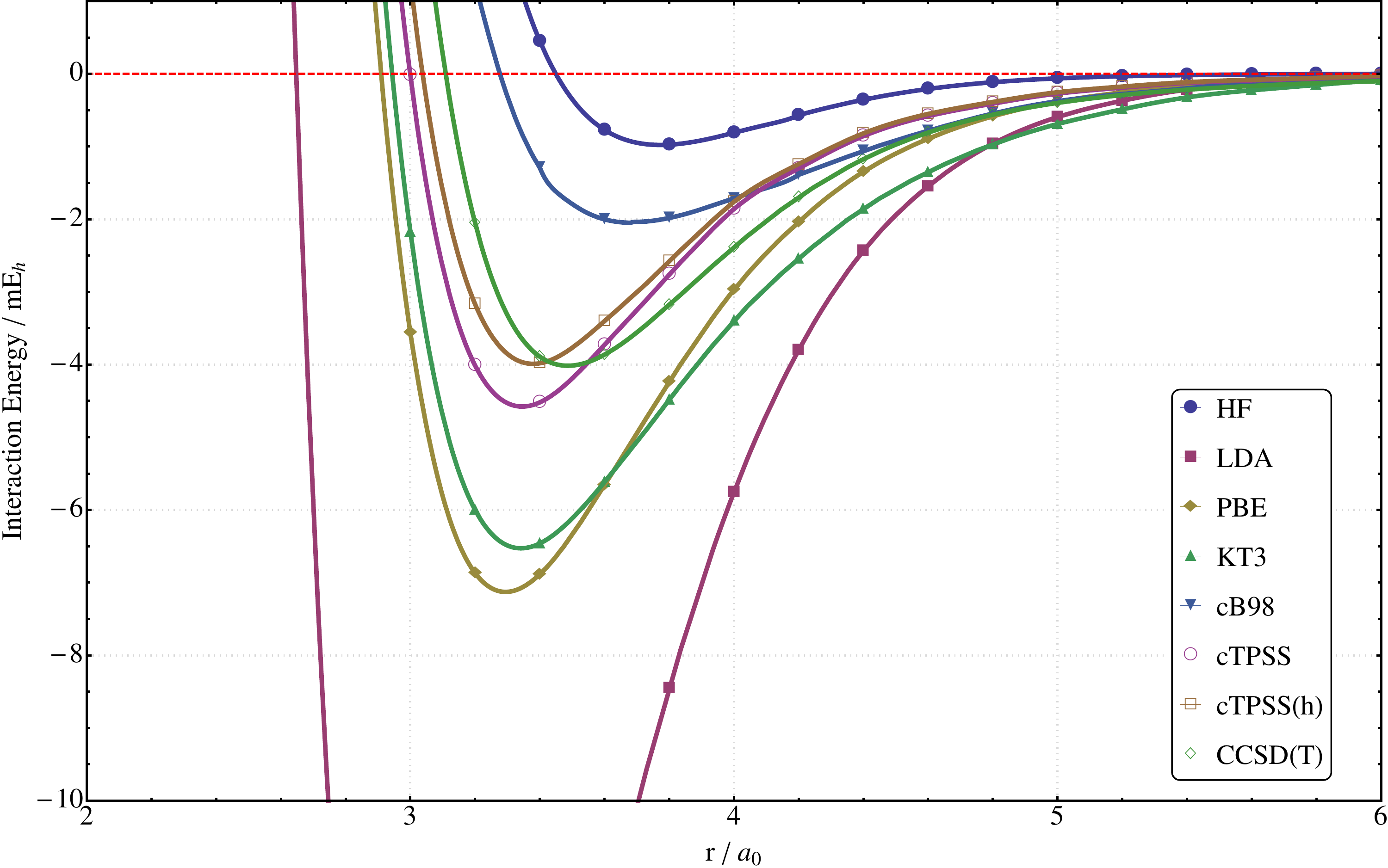}
\caption{Potential energy curve for the lowest $^1\Sigma_g^+$ state of Ne$_2$ in a magnetic field of 1 a.u. perpendicular to the bonding axis for a variety of methods with the aug-cc-pCVTZ basis set.}\label{fig:nene}
\end{figure}

To investigate the performance of cB98, cTPSS and cTPSS(h) in strong magnetic fields potential energy profiles were calculated for H$_2$, He$_2$, NeHe and Ne$_2$. In particular, we consider the $^3\Sigma_u^+ (1 \sigma_g 1 \sigma_u^*)$ state of H$_2$ and the lowest $^1\Sigma_g^+$ states of He$_2$, HeNe and Ne$_2$. Each of these states is repulsive or weakly dispersion bound in the absence of a magnetic field but become more strongly bound when a field is applied. We note that only the $^3\Sigma_u^+ (1 \sigma_g 1 \sigma_u^*)$ of H$_2$ is an overall ground state in the presence of the field. These states were compared against results from accurate CCSD(T) potential energy curves calculated using a recent non-perturbative implementation by Stopkowicz et al\cite{CC_IMPLEMENTATION} in the \london program. For comparison we have also generated similar profiles with standard LDA and GGA density functionals as well as with HF theory. The calculated potential energy curves for H$_2$, He$_2$, NeHe and Ne$_2$ are shown in Figures \ref{fig:h2}, \ref{fig:hehe}, \ref{fig:nehe} and \ref{fig:nene}, respectively. Equilibrium bond lengths and dissociation energies were determined numerically and are presented in Table \ref{tab:combined}. 
\begin{table}

\caption{Dissociation energies and equilibrium bond lengths for H$_2$ and rare gas dimers in a 1 a.u. magnetic field perpendicular to the internuclear axis.}
\label{tab:combined}
\begin{tabular}{l|cccc|cccc}
\hline
\hline
          	& \multicolumn{4}{c|}{$R_\text{e}$ / $a_0$} & \multicolumn{4}{c}{$D_e$ / mE$_\text{h}$} \\
         	& H$_2$ 	& He$_2$ & NeHe 	& Ne$_2$ & H$_2$ 	& He$_2$	& NeHe   	& Ne$_2$ \\
 \hline                                                      
HF        	&  2.708 	& 3.296   & 3.543 	& 3.773   	& 2.340  	& 0.218 	& 0.482  	& 0.978  \\
LDA      	&  2.374  	& 2.550   & 2.846 	& 3.062   	& 14.81  	& 8.231 	& 11.817 	& 19.730 \\
PBE      	&  2.514  	& 2.810   & 3.080 	& 3.294   	& 5.985  	& 2.463 	& 3.938  	& 7.128  \\
KT3      	&   2.511 	& 2.852   & 3.121 	& 3.342   	& 4.637  	& 2.661 	& 3.937  	& 6.527  \\
cB98     	&   2.640 	& 3.203   & 3.472 	& 3.676   	& 1.328  	& 1.011 	& 1.514  	& 2.050  \\
cTPSS  	&   2.564  	& 2.864   & 3.124 	& 3.346   	& 5.255  	& 1.307 	& 2.344  	& 4.577  \\
cTPSSh  	&    2.558 	& 2.879   & 3.154 	& 3.379   	& 5.263  	& 1.245 	& 2.130  	& 3.990  \\
CCSD(T)  &   2.578 	& 2.977   & 3.248 	& 3.487   	&  4.554 	& 1.259 	& 2.217 	& 4.016  \\
FCI$^{a}$ &  2.578 	& 2.975   & -    		& -      	&  4.554 	& 1.271 	& -      	& -      \\
\hline
\hline
\end{tabular}\\
\footnotesize{
$^{a}$ The He$_2$ FCI calculations use the aug-cc-pVTZ basis set.}
\end{table}

The $^3\Sigma_u^+ (1 \sigma_g 1 \sigma_u^*)$ state of the H$_2$ molecule in a perpendicular field was examined at the FCI level by Lange \emph{et al.}~\cite{Lange2012}. The potential energy curves for this state are shown in Figure~\ref{fig:h2}. HF strongly underbinds this state in comparison with the FCI data. In contrast LDA strongly overbinds, a tendency which is largely corrected by the PBE functional and further improved by the cTPSS and cTPSS(h) models. These trends are reflected in the equilibrium bond lengths and dissociation energies in Table~\ref{tab:combined}. Although not highly accurate the cTPSS and cTPSS(h) models give a reasonable qualitative description of the potential energy curve. The empirically parameterized KT3 functional is interesting because it gives simultaneously a reasonable estimate of both the equilibrium bond length and dissociation energy. However, at intermediate separation an unphysical barrier is observed. For B98 an even more pronounced barrier is present and the potential energy curve is generally even less accurate than Hartree--Fock theory. This may suggested that heavily parameterized functional forms, determined to perform well at zero field, may not be the best candidates for use in strong-field CDFT studies.  

Examining the potential energy curves for He$_2$ in Figure~\ref{fig:hehe} we see that HF tends to under-bind with a bond length of $3.30$ $a_0$ compared with the CCSD(T) value of $2.98$ $a_0$. Similarly the HF dissociation energy of $0.218$ mE$_\text{h}$ is much smaller than the corresponding CCSD(T) value of $1.259$ mE$_\text{h}$. For this small system we were able to compare the CCSD(T) results with FCI values (calculated in the slightly smaller aug-cc-pVTZ basis), as expected the agreement is excellent -- the corresponding potential energy curves are essentially indistinguishable on the scale of Figure~\ref{fig:hehe}. For LDA we see a strong tendency to overbind giving much too short $R_\text{e}$ values and much too large estimates of $D_e$. The GGA functionals PBE and KT3 show considerable improvement over LDA, however, they still strongly overbind. The improvement for the mGGA functionals is striking -- in particular TPSS and TPSS(h) give a good qualitative description of the potential energy curve. The corresponding $R_\text{e}$ and $D_\text{e}$ values indicate that there still remains a tendency towards over binding but this is greatly reduced. 

The cB98 functional tends to show more significant under binding. Here we note that the arguments used in the construction of cB98 and cTPSS are rather different. In particular, cB98 is an empirically parameterized functional (see the appendix), whereas cTPSS is constructed based on the satisfaction of known exact conditions. In this work we have used the parameters determined in Ref.~\onlinecite{Becke1998} to define the cB98 form. These parameters were determined at zero field and from post-LDA calculations -- as a result they may not be optimal for fully self-consistent calculations in the presence of a magnetic field. On the other hand the cTPSS functional is designed to satisfy selected constraints at zero field and it could be argued that in the presence of a field both the B98 and TPSS based functionals are open to further optimization, though this is beyond the scope of the present work.  

The stability of the mGGA functionals is particularly evident in the the strong field regime when one compares the present results for He$_2$ with those for the VRG-based estimates in Figure 7 of Ref.~\onlinecite{Tellgren2014}. The VRG approaches led to very difficult SCF convergence and complex potential energy curves with a strong unphysical over binding. The mGGAs considered here are un-problematic in practical application and yield results surprisingly close to the CCSD(T) estimates.

Similar qualitative trends are observed for the NeHe and NeNe dimers in Figures~\ref{fig:nehe} and~\ref{fig:nene}. Again LDA and GGA functionals are not sufficiently accurate for practical use and the mGGA functionals provide a large improvement. The cTPSS and cTPSS(h) results remain impressive -- with cTPSS(h) being consistently slightly more accurate than cTPSS. This trend is reflected in both the potential energy curves and the $R_\text{e}$ and $D_\text{e}$ values in Table~\ref{tab:combined}.

\section{Interpretation of paramagnetic bonding in the KS-CDFT framework}\label{sec:KSCDFTpbond}
The mGGA CDFT functionals offer a computationally cheap correlated method for the examination of the exotic bonding mechanisms observed in a strong magnetic field. In many areas of chemistry the nature of bonding, chemical reactions, spectra, and properties of molecular species are interpreted qualitatively in terms of orbital interactions. We now consider the extent to which information from KS-CDFT calculations can aid in simple interpretation of the perpendicular paramagnetic bonding interactions. 

We begin by considering a molecular orbital analysis of the perpendicular paramagnetic bonding. KS-CDFT calculations provide a simple set of canonical molecular orbitals, which can be used to construct the electronic density via
\begin{equation}\label{KSRHO}
\rho(\mathbf{r}) = \sum_i | \varphi_i^{}(\mathbf{r}) \varphi_i^*(\mathbf{r}) |^2 = \sum_{\gamma \zeta} \omega_\gamma(\mathbf{r}) D_{\gamma\zeta} \omega_{\zeta}^*(\mathbf{r})
\end{equation}
Here we note that the occupied KS orbitals $\varphi_i^{}(\mathbf{r})$ can be complex in the presence of a magnetic field. In the second equality the density is expressed in terms of the one-particle density matrix $D_{\gamma\zeta}$ and the basis functions $\omega_\gamma(\mathbf{r})$. We will commence by considering how the molecular orbital energies associated with H$_2$ and the rare gas dimers change upon application of a magnetic field as the perpendicular paramagnetic bonding in Section~\ref{sec:KSMO} evolves. Since the orbitals themselves can be complex in the presence of a field we then proceed in Section~\ref{sec:KSRHO} to analyze the bonding in terms of the changes in electronic density of Eq.~(\ref{KSRHO}) as a function of field, which is naturally a real observable quantity. 
 
 \subsection{KS-CDFT molecular orbital analysis}\label{sec:KSMO}
KS molecular orbitals have been widely used as an interpretive aid in chemical applications throughout the literature. The KS orbitals are defined to minimize the non-interacting kinetic energy and yield the physical electronic density via Eq.~(\ref{KSRHO}). They also have appealing properties; for example the highest occupied MO energy is minus the first ionization potential (IP)~\cite{Perdew1982,Almbladh1985} and the remaining orbital energies can be interpreted as Koopman's type approximations to higher IPs~\cite{Bartlett2005}. The extent to which these properties hold for general practical approximations has been a subject of debate in the literature~\cite{Baerends1997a,Stowasser1999}, as has the interpretation of KS virtual orbitals~\cite{Meer2014} owing to the role of the integer discontinuity~\cite{Perdew1982}, which is missing from common approximations. However, from a practical standpoint it is widely accepted that interpretations based on occupied KS orbitals (to which we limit the following discussion) are theoretically justified and their utility has been borne out in many practical applications.    

\begin{figure}[h]
\includegraphics[width=\columnwidth]{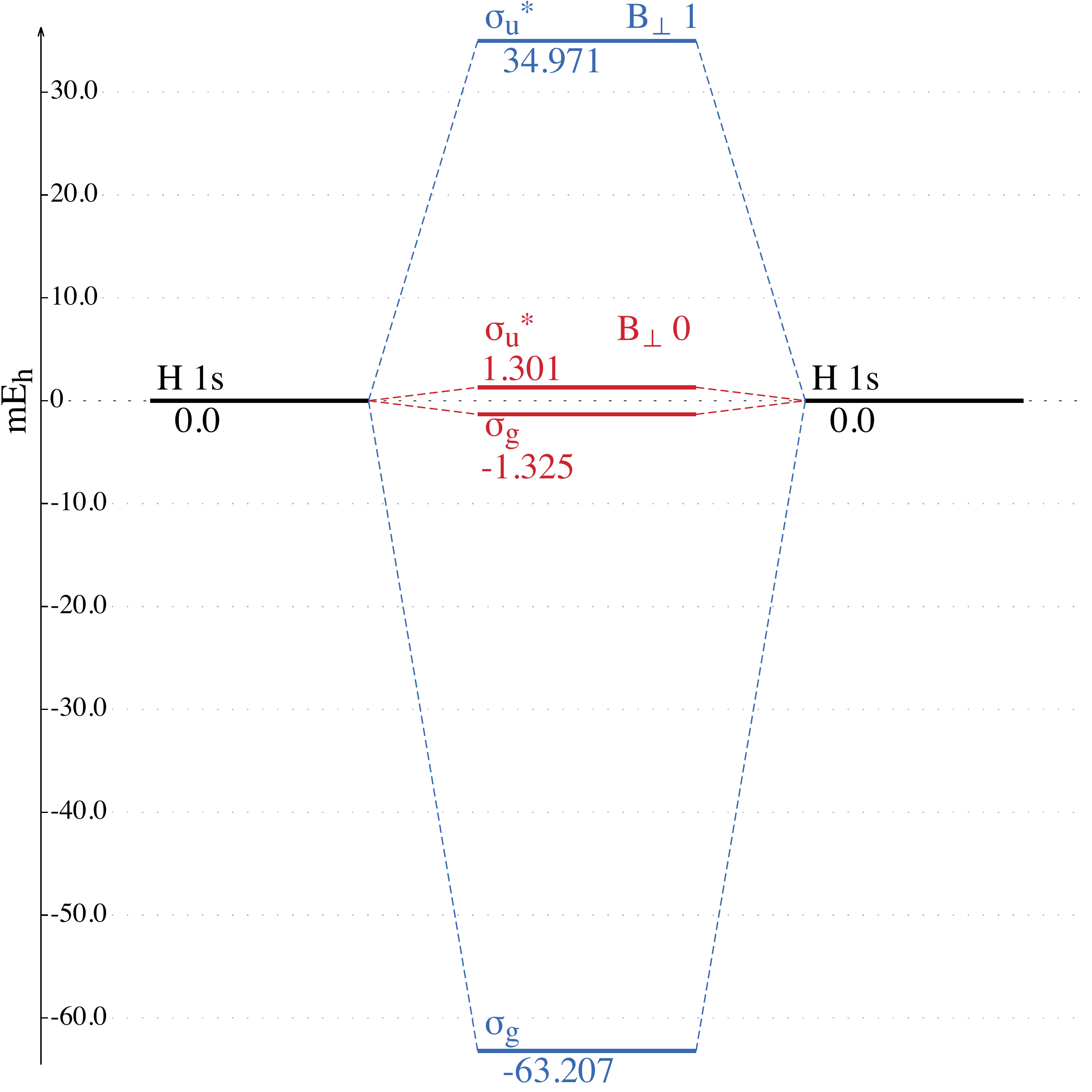}
\caption{Relative orbital energies for the occupied $\sigma_g$ and $\sigma_u^*$ CDFT molecular orbitals for H$_2$ relative to the atomic 1s orbitals, with (blue) and without (red) a field of 1.0 a.u. perpendicular to the interatomic axis.}\label{fig:H2MOs}
\end{figure}

\begin{figure}[h]
\includegraphics[width=\columnwidth]{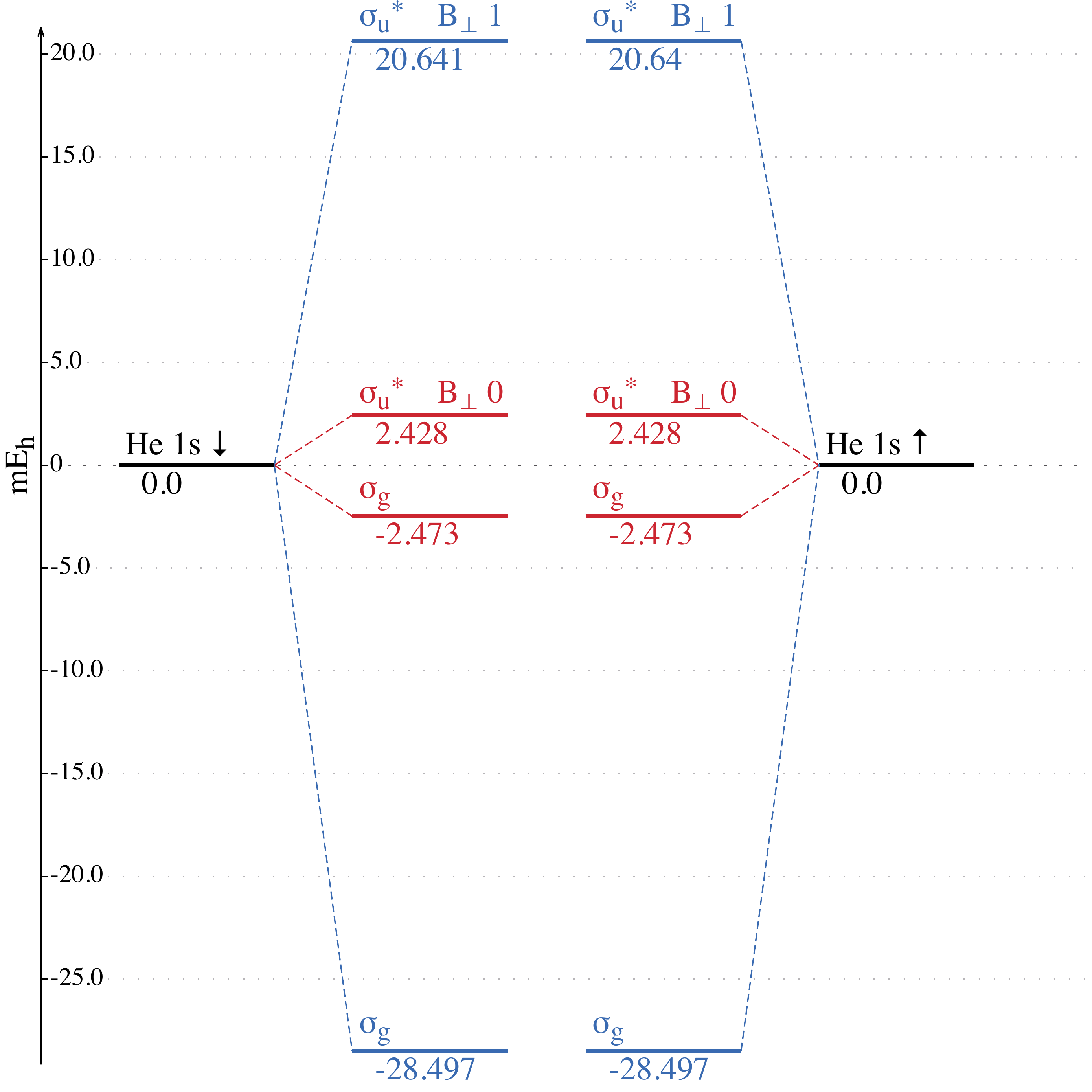}
\caption{Relative orbital energies for the occupied $\sigma_g$ and $\sigma_u^*$ CDFT molecular orbitals for He$_2$ relative to the atomic 1s orbitals, with (blue) and without (red) a field of 1.0 a.u. perpendicular to the interatomic axis.}\label{fig:HeMOs}
\end{figure}

We now consider how the KS orbital energies change upon application of a perpendicular magnetic field of 1 a.u. Given that the cTPSS based models seem to be the most reliable of those studied in the present work we consider how the orbital energies from this functional change in Figures~\ref{fig:H2MOs} and~\ref{fig:HeMOs} for H$_2$ and He$_2$, respectively. For the H$_2$ molecule in the $^3\Sigma_u^+ (1 \sigma_g 1 \sigma_u^*)$ state we consider the energy of the occupied  $\sigma_g$ and $\sigma_u^*$ orbitals change upon bonding. In particular, we plot orbital energies in the absence of a field and in a perpendicular field of 1 a.u. relative to those of the atomic orbitals in the same field. We have plotted the orbital energies at an internuclear separation $R_e = 2.564$ a.u. consistent with the cTPSS equilibrium bond length in the presence of a field.

We see that in the absence of a field the singly occupied $\sigma_g$ orbital is stabilized by $1.33$ mE$_\text{h}$, whilst the singly occupied $\sigma_u^*$ is destabilized by $1.30$ mE$_\text{h}$ relative to the 1s hydrogen orbitals. This is consistent with a net bond order of zero and a repulsive profile for the corresponding potential energy curve. In the perpendicular field of 1 a.u. we see that, relative to the hydrogen 1s orbital in the same field, the $\sigma_g$ orbital is stabilized by $63.21$ mE$_\text{h}$, whilst the $\sigma_u^*$ orbital is destabilized by $34.97$ mE$_\text{h}$. This greater stabilization of the $\sigma_g$ orbital leads to a net bonding interaction, consistent with the analysis of Ref.~\onlinecite{Lange2012}. This illustrates that the KS orbitals can be a useful tool in rationalizing this exotic bonding phenomenon. 

A similar analysis can be carried out for the lowest $^1\Sigma_g^+$ state of He$_2$ and is presented in Figure~\ref{fig:HeMOs}. In this plot we have separated the spin down and spin up orbitals and defined their energies relative to atomic orbitals of the same spin in the same field. Defined in this way the offset between the orbitals of different spin due to the Zeeman interaction is removed from the plot. Again the energies correspond to the cTPSS He$_2$ equilibrium internuclear separation $R_e = 2.864$ a.u. in the presence of a perpendicular field. In the absence of a field we see that again the relative stabilization of $\sigma_g$ and destabilization of $\sigma_u^*$ are approximately compensatory, whilst in the presence of a field the $\sigma_g$ orbital is more stabilized than the $\sigma_u^*$ orbital is destabilized. It is also clear that the extra stabilization of the $\sigma_g$ orbital of $\sim 8$ mE$_\text{h}$ is considerably less than the $\sim 28$ mE$_\text{h}$ observed for H$_2$. This is consistent with the strength of binding exhibited for these species in Figures~\ref{fig:h2} and \ref{fig:hehe} as well as in Table~\ref{tab:combined}.  

Similar orbital energy diagrams may be constructed for the HeNe and NeNe systems, however, they become significantly more complex due to large differences between the orbital energies in HeNe and the splitting of the p-orbitals in NeNe. We therefore consider an alternative visualization of the bonding effects in these systems based on the charge and (physical) current density differences.  

\subsection{Electron density analysis}\label{sec:KSRHO}
For each of the species He$_2$, HeNe and Ne$_2$, we have performed calculations of the electronic density and current density in the presence of varying perpendicular magnetic fields ($B_{\perp}$) using the cTPSS functional at the corresponding equilibrium geometries. Here we consider the density change $\Delta \rho_{B_{\perp}}(\mathbf{r})$ for each system relative to the isolated atoms in the same field,
\begin{equation}\label{rhodiff}
\Delta \rho_{B_{\perp}}(\mathbf{r}) = \rho^{\text{Dimer}}_{{B}_{\perp}}(\mathbf{r}) - \sum_{i=1}^{N_{\text{atoms}}} \rho^{i}_{B_{\perp}}(\mathbf{r})
\end{equation}
This difference density allows for a visualization of the nature of the paramagnetic bonding in these systems. In a similar manner one can consider the (gauge invariant) \emph{physical} current density difference $\Delta \mathbf{j}$ 
\begin{equation}\label{jdiff}
\Delta \mathbf{j}_{B_{\perp}}(\mathbf{r}) = \mathbf{j}^{\text{Dimer}}_{{B}_{\perp}}(\mathbf{r}) - \sum_{i=1}^{N_{\text{atoms}}} \mathbf{j}^{i}_{B_{\perp}}(\mathbf{r})
\end{equation}

\begin{figure*}
\includegraphics[width=\textwidth]{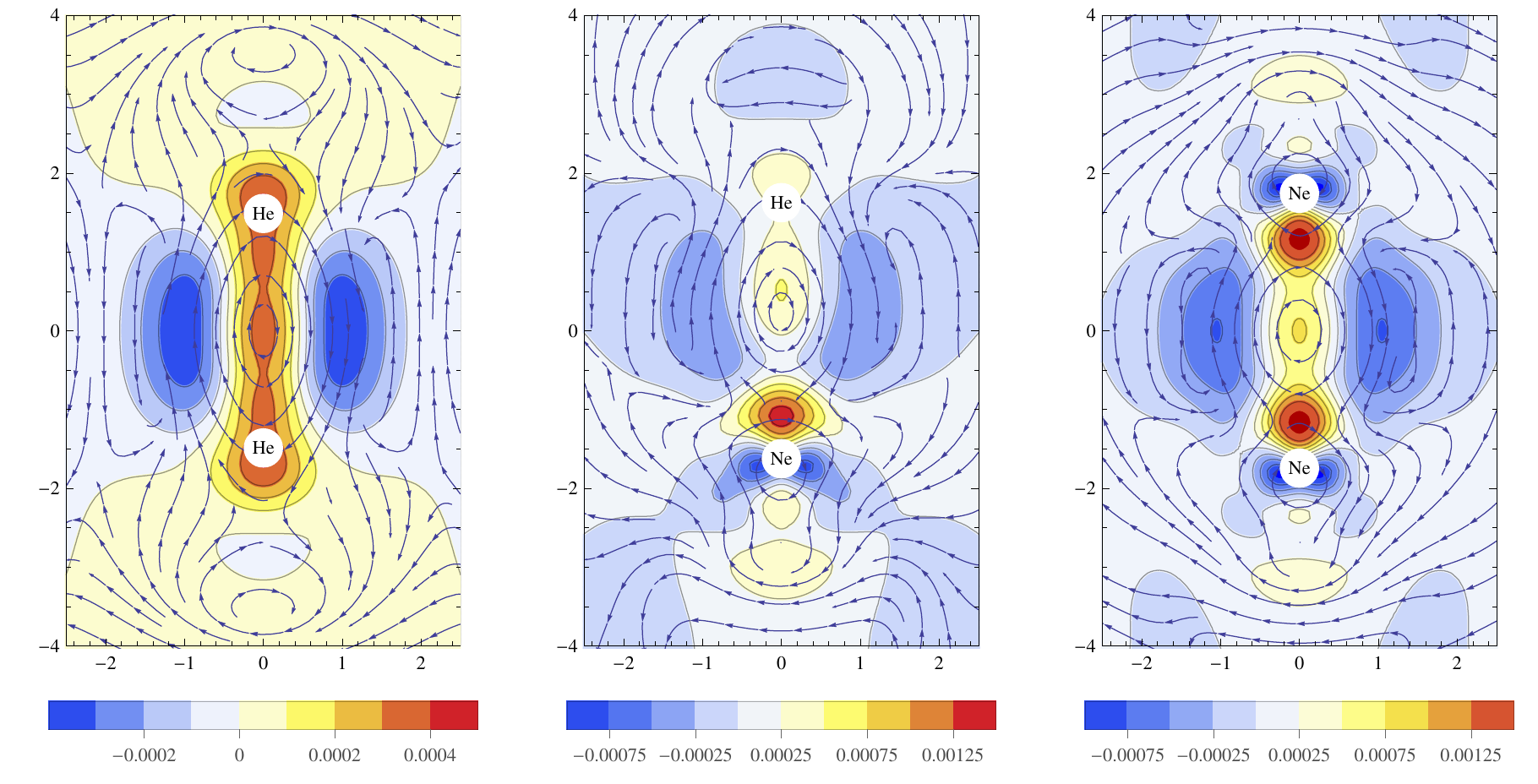}
\caption{$\Delta \rho_{B_{\perp}}(\mathbf{r})$ (coloured contours) and $\Delta \mathbf{j}_{B_{\perp}}(\mathbf{r})$ (streamlines) for the rare gas dimers He$_2$ (left), HeNe (middle) and Ne$_2$ (right) in a ${B}_{\perp} = 1.0$ a.u. magnetic field. The internuclear axis is aligned with the $z$-axis and the plots are show in the $yz$-plane intersecting the atomic positions.}\label{fig:densDiff}
\end{figure*}

In Figure \ref{fig:densDiff} we present plots of the density differences in Eqs.~(\ref{rhodiff}) and (\ref{jdiff}) for each of the species with ${B}_{\perp} = 1.0$ a.u.
The shading of the contours represents the buildup (red) or depletion of the charge density (blue), relative to two non-interacting atoms in the same field. In all three cases there is a clear build up of density between the atoms consistent with bonding. The charge density difference is elongated along the field, above and below the plane of the plots. The streamlines show the vector field associated with the current density difference. Paratropic circulations are clearly visible over the centre of the bonds where charge density accumulates, and diatropic circulations are visible in regions where the charge density is depleted. We have confirmed that for higher fields the alignment between para- / dia-tropic circulations and charge accumulation / depletion becomes more pronounced.

This picture of the perpendicular paramagnetic bonding suggests that as the charge density is elongated along the field the constituent atoms may approach one another more closely. As they do so they experience a greater nuclear-nuclear repulsion, which may be screened by a rearrangement of the charge density towards the bond centre. This rearrangement is accompanied by consistent para- and dia-tropic current circulations. The cTPSS functional used in this work provides a simple, computationally cheap, route to perform analysis of the bonding encountered in the strong field regime.  

\section{Conclusion}\label{sec:conclusions}

In this work we have implemented a previously detailed\cite{Dobson1992, Dobson1993, Tao2005a} modification to the kinetic energy density term in mGGA functionals to perform non-perturbative cDFT calculations in a stable, gauge invariant manner using the \london program~\cite{Tellgren2008}. The modified mGGA functionals cB98, cTPSS and cTPSS(h) show a level of accuracy in predicting weak field magnetic properties that is competitive with existing GGA functionals without any additional fitting. The functionals cTPSS and cTPSS(h) show excellent prediction of perpendicular paramagnetic bonding behaviour, suggesting that the modification of the kinetic energy density is a viable route for the incorporation of current effects in standard mGGA density-functionals. In contrast to vorticity dependent forms previously studied~\cite{Tellgren2014,Zhu2014} the functionals exhibited excellent numerical stability in the finite field setting without the need for delicate numerical regularizations. 

Whilst the mGGA results show considerable promise in the high field regime, their performance in the weak field regime is perhaps disappointing -- leading to only modest improvements of conventional GGA forms. To some extent this may be due to the fact that mGGAs contain many of the shortcomings associated with GGA forms. For example, the functionals still have potentials with incorrect asymptotic behaviour -- particularly for the exchange contribution. Since the current effects at weak field are not dominant but rather add small corrections to the predominantly Coulombic exchange and correlation interactions then it may be necessary to further improve the underlying functional forms before the true impact of the current terms can be assessed in this regime.

We expect that this approach to include current dependence should play a central role in the future development of new CDFT functionals and many avenues are open for development. Obvious possibilities include the generalization of range-separated mGGA functionals~\cite{Goll2009,Peverati2011} to obtain a better balance of errors between exchange, correlation and current contributions and the re-parameterization of functionals either empirically or via the consideration of alternative exact conditions in their construction. In the latter category we note that the presence of a magnetic field causes compression of the electronic density in two dimensions perpendicular to the field and elongation along it. As a result non-uniform coordinate scaling relations may be one example of conditions that may provide a powerful tool in further development. The \london program~\cite{Tellgren2008} provides a powerful platform for this work since CDFT approaches can be calibrated against accurate \textit{ab initio} data for a range of field strengths, where experimental data may be scarce. In the future we hope that these relatively inexpensive CDFT approaches may then be applied to the study of larger systems such as those in Refs.~\onlinecite{Murdin2013,Litvinenko2014}.   

\appendix
\section{meta-GGAs in CDFT}
We present the key working equations defining the B98 and TPSS functionals, indicating how the the modified $\tilde \tau_{\sigma}$ of Eq.~(\ref{tauGI}) enters each functional. 
\subsection{cB98}
The B98 functional~\cite{Becke1998} has a general construction that is similar to the popular B97 functional form~\cite{Becke1997}, however, instead of using reduced spin-density gradients it makes use of an inhomogeneity parameter $q_\sigma$
\begin{equation}
q_\sigma = \frac{(Q_\sigma-Q_\sigma^{\text{UEG}})}{|Q_\sigma^{\text{UEG}}|}
\end{equation} 
where the exchange hole curvature for the uniform electron gas takes the simple form
\begin{equation}
Q_\sigma^{\text{UEG}} = -\frac{1}{5} (6 \pi^2)^{2/3} \rho_\sigma^{5/3}
\end{equation}
and $Q_\sigma$ is defined in Eq.~(\ref{Q}).
This inhomogeneity factor, $q_\sigma$, controls the enhancement or attenuation of the exchange and correlation energy over the uniform gas values. 

The exchange component of the functional takes the form
\begin{align}
\label{B98x}
E_{\text{x},\sigma} = \int e_{\text{x},\sigma}^{\text{UEG}}(\rho_\sigma) g_{\text{x},\sigma}(q_\sigma) \mathrm{d} \mathbf{r}
\end{align}
and the opposite- and same-spin components of the correlation energy take the forms
\begin{equation}
\label{B98cab}
E_{\text{c},\alpha \beta} = \int e_{\text{c},\alpha \beta}^{\text{UEG}}(\rho_\alpha, \rho_\beta) g_{\text{c},\alpha \beta}(q_\text{avg}) \mathrm{d} \mathbf{r}
\end{equation}
\begin{equation}
\label{B98caa}
E_{\text{c},\sigma \sigma} = \int e_{\text{c},\sigma \sigma}^{\text{UEG}}(\rho_\sigma) f_\sigma^{\text{SCC}} g_{\text{c},\sigma \sigma}(q_\sigma) \mathrm{d} \mathbf{r}.
\end{equation}
The functions $g_{\text{x},\sigma}(q_\sigma)$, $g_{\text{c},\alpha \beta}(q_\text{avg})$ and $g_{\text{c},\sigma \sigma}(q_\sigma)$ are dimensionless inhomogeneity correction factors depending on $q_\sigma$ and $q_\text{avg} = \frac{1}{2}(q_\alpha + q_\beta)$. In addition the same-spin correlation energy contains a self-correlation correction (SCC) factor
\begin{equation}
f_\sigma^{\text{SCC}} = \left[  \tau_\sigma - \frac{1}{4} \frac{(\nabla \rho_\sigma)^2}{\rho_\sigma}  \right] / \tau_\sigma
\end{equation}
This factor varies between $0$ and $1$ and vanishes in one-orbital regions, ensuring the functional is self-correlation free. For convenience in deriving fitted forms for the inhomogeneity correction factors $g$, Becke proposed the following transformation to a finite interval,
\begin{equation}
w = \frac{\gamma q}{\sqrt{1+ \gamma^2 q^2}}
\end{equation}
where $\gamma$ is a parameter to be determined and separate transformations are carried out for the exchange, opposite-spin correlation and like-spin correlation respectively using the appropriate definitions of $q$ as in Eqs~(\ref{B98x})--(\ref{B98cab}). Based on calculations of atomic exchange--correlation energies Becke proposed the values $\gamma_{\text{x},\sigma}=0.11$, $\gamma_{\text{c},\alpha \beta}=0.14$ and $\gamma_{\text{c},\sigma \sigma}=0.16$. The inhomogeneity corrections $g$ were then determined by fitting a power series expansion of the form
\begin{equation}
g = \sum_{i=0}^m c_i w^i
\end{equation}
with $m=2$ chosen to prevent unphysical over-fitting. This fitting was carried out using the G2 thermochemical dataset and basis set free, post-LSDA, calculations. The optimal parameters can be found in Ref.\citenum{Becke1998}. 

Here we use this parameterization directly but note that in future these parameters could be re-optimized based on self-consistent data. In addition an amount of Hartree--Fock exchange is included with weight $c_{\text{x}}=0.1985$. The resulting B98 functional therefore possesses a high degree of non-locality and may be classified as a hybrid mGGA functional with dependence not only on $\tau$ but also on the laplacian of the density $\nabla^2 \rho$.

The original definition of B98 utilized the zero-field exchange hole curvature of Eq.~(\ref{Q}) in its definition. However, it was noted~\cite{Becke1998} that this form can be readily extended to include current effects via Eq.~(\ref{cQ}) and it is this avenue that we explore in the present work. Unless otherwise stated we employ this modified form throughout and denote it as cB98. 

\subsection{cTPSS}
One of the most widely used meta-GGA functionals is due to Tao, Perdew, Staroverov and Scuseria (TPSS)\cite{Tao2003}. This functional is designed to satisfy exact constraints without empirical parameters and as such is an interesting candidate to study in the context of generalization to finite magnetic field strengths where much less is known about the performance of approximate functionals. In this functional the ratio 
\begin{equation}\label{zTPSS}
z = 2 \tau^\text{W}/\tau, \hspace{0.2in} \tau = \sum_\sigma \tau_\sigma, \hspace{0.2in} \tau^{\text{w}} = \frac{1}{8} \frac{|\nabla \rho|^2}{\rho} 
\end{equation}
plays a key role as a dimensionless inhomogeneity parameter, along with 
\begin{equation}
p = \frac{|\nabla \rho|^2}{4 (3\pi^2)^{2/3} \rho^{8/3}} = s^2
\end{equation}
Note that throughout this work we use the definition of $\tau$ in Eq.~(\ref{tau}), which does not include the factor of 1/2 commonly employed. The exchange functional then takes the form
\begin{equation}
E_{\text{x}}[\rho] = \int \rho \varepsilon_{\text{x}}^{\text{UEG}}(\rho) F_{\text{x}}(p,z)
\end{equation}
where the precise details of the form chosen for the enhancement factor $F_{\text{x}}(p,z)$ can be found in Ref.~\citenum{Tao2003}. The correlation energy takes the form
\begin{align}\label{cTPSSc}
E_{\text{c}}&[\rho_\alpha, \rho_\beta, \nabla \rho_\alpha,\nabla \rho_\beta, \tau] = \nonumber\\
 &\int \rho \varepsilon_{\text{c}}^{\text{revPKZB}}(\rho_\alpha, \rho_\beta, \nabla \rho_\alpha,\nabla \rho_\beta, \tau) \nonumber \\
&\times \left[1 + d \varepsilon_{\text{c}}^{\text{revPKZB}}(\rho_\alpha, \rho_\beta, \nabla \rho_\alpha,\nabla \rho_\beta, \tau) (\tau^{\text{W}}/\tau)^3 \right] \mathrm{d}\mathbf{r} 
\end{align}
where $d = 2.8$ hartree$^{-1}$.

Using the replacement in Eq.~(\ref{tauGI}) leads to modifications in the exchange contribution via $z$ in Eq.~(\ref{zTPSS}) and in the correlation energy as shown in Eq.~(\ref{cTPSSc}). This modified form is noted cTPSS and is consistent with that used in the response implementation of Ref.~\onlinecite{Bates2012}.

\begin{acknowledgements}
A. M. T. is grateful for support from the Royal Society University Research Fellowship scheme. We are grateful for access to the University of Nottingham High Performance Computing Facility. 
   This work was supported by the Norwegian Research Council through
   the CoE Centre for Theoretical and Computational Chemistry (CTCC)
   Grant No.\ 179568/V30 and the Grant No.\ 171185/V30 and through the
   European Research Council under the European Union Seventh Framework
   Program through the Advanced Grant ABACUS, ERC Grant Agreement No.\
   267683. 
\end{acknowledgements}
\end{document}